# Human decisions in moral dilemmas are largely described by Utilitarianism: virtual car driving study provides guidelines for ADVs

Anja Faulhaber[1,#], Anke Dittmer[1,#], Felix Blind[1,#], Maximilian A. Wächter[1,#*], Silja Timm[1,#], Leon R. Sütfeld[1], Achim Stephan[1], Gordon Pipa[1], Peter König[1,2]

[1] Institute of Cognitive Science, University of Osnabrück, Osnabrück, Germany
[2] Department of Neurophysiology and Pathophysiology, Center of Experimental Medicine, University Medical Center Hamburg-Eppendorf, Hamburg, Germany

[#] Shared First Authorship

*Correspondence: Maximilian A. Wächter. Institute of Cognitive Science, Wachsbleiche 27, University of Osnabrück, D-49090 Osnabrück, e-mail: mwaechter@uni-osnabrueck.de

## Abstract

Ethical thought experiments such as the trolley dilemma have been investigated extensively in the past, showing that humans act in a utilitarian way, trying to cause as little overall damage as possible. These trolley dilemmas have gained renewed attention over the past years; especially due to the necessity of implementing moral decisions in autonomous driving vehicles (ADVs). We conducted a set of experiments in which participants experienced modified trolley dilemmas as the driver in a virtual reality environment. Participants had to make decisionsbetween two discrete options: driving on one of two lanes where different obstacles came into view. Obstacles included a variety of human-like avatars of different ages and group sizes. Furthermore, we tested the influence of a sidewalk as a potential safe harbor and a condition implicating a self-sacrifice. Results showed that subjects, in general, decided in a utilitarian manner, sparing the highest number of avatars possible with a limited influence of the other variables. Our findings support that people's behavior is in line with the utilitarian approach to moral decision making. This may serve as a guideline for the implementation of moral decisions in ADVs.

# Introduction

Since their invention in the 19th century, cars have considerably influenced the townscapes and societies all over the world. Due to the continuous development and increasing sophistication of vehicles, this impact is still ongoing. It even seems that we are getting closer reaching another milestone: a car thatis capable of driving without a human driver. In the last years, there has been a considerable advance in the development of such autonomous vehicles. Many features of automation, such a cruise control, and camera based blind spot assistance and parallel parking have already become a standard in modern cars. The majority of car manufacturers, as well as service providers like Uber, are currently working on ADV's and planning to commercially market them by latest 2025 (Harst, 2016). However, with the development of disruptive technologies, new problems arise. As the introduction of autonomous vehicles might have a large impact on society, critical issues spread over a wide range of areas including psychological, and ethical, socioeconomic as well as legal aspects. These far-reaching implications are also evident in the fact that autonomous cars have recently become a topic of interest in science. In particular, the implementation of moral decisions in self-driving cars remains widely debated (e.g.Bonnefon, Shariff, & Rahwan, 2016; Hevelke & Nida-Rümelin, 2014, 2015a, 2015b).

Moral decisions by autonomous systems are often discussed on the basis of trolley dilemmas. The classical trolley dilemma was first introduced in 1967 as a philosophical thought experiment (Foot, 1967). The key element is a trolley that is heading straight towards a group of people, e.g. five, on the rails unable to escape. There is, however, a sidetrack on which a single person, who is unaware of the trolley, stands. The participant in this thought experiment is standing next to a lever that enables the trolley to switch to the sidetrack. Without intervention, the trolley will kill the five people on the main track. On pulling the lever, the trolley will continue on the sidetrack killing only one person, resulting in a moral dilemma for the participant. How do people decide in such situations and which moral principles govern their decision process? This question has been investigated and debated extensively since (e.g. Mikhail, 2007; Thomson, 1976, 1985; Unger, 1996). So far, research on modified trolley dilemmas has shown that people in general act utilitarian and are relatively comfortable with utilitarian ADVs, programmed to minimize harm (Bonnefon et al., 2016; Skulmowski, Bunge, Kaspar, & Pipa, 2014). In contrast, German law interdicts the evaluation of human life (Art. 1 Abs. 1 GG) in any way. Thus, the trolley dilemma as applying to autonomous systems is still not resolved.

Moreover, studies including such trolley dilemmas were traditionally carried out in the form of philosophical essays. This means that the material was presented to the participants in the form of written scenario descriptions, sometimes with additional pictorial representations. This way of presenting the dilemma introduces issues like the disregard of important contextual and situational influences in moral decision-making(Skulmowski et al., 2014). New technologies such as virtual reality (VR) help to remedy these insufficiencies. In this context, trolley dilemmas have recently experienced a revival in science(e.g. Navarrete, McDonald, Mott, & Asher, 2012; Pan, Banakou, & Slater, 2011; Patil, Cogoni, Zangrando, Chittaro, & Silani, 2014; Skulmowski et al., 2014). The immersion thatVR environments provide serves to improve ecological validity while maintaining control over experimental variables (Madary & Metzinger, 2016). In consequence, it seems that we should also benefit from VR technologies to find out how people actually behave in dilemma situations during car driving when they are immersed in a more realistic experimental environment as opposed to their decisions in plain thought experiments.

Furthermore, many possible modifications of the trolley dilemma elicit open questions. For example, different characteristics of the potential victims might influence the human decision process. Previous studies have shown that children were saved more often than adults, so that age of the potential victims might play a role in decision-making(Sütfeld, Gast, König, & Pipa, 2016). In the context of autonomous cars, there are additionally certain traffic-specific aspects worth considering. German law, for instance, inflicts punishment to prevent citizens from driving on the sidewalk in order to provide a safe space for pedestrians in traffic (STVO §49 Abs.2). This might lead to an internalized reluctance to drive on the sidewalk that might also influence the decision process in a trolley dilemma modified accordingly. Furthermore, there are possible scenarios in which people can only save lives by sacrificing their own. Despite evidence from surveys that revealed a willingness to use self-sacrificing ADVs (Bonnefon et al., 2016), it is questionable whether people would indeed act this way in a realistic setting.

The present study addresses these open questions and improves the experimental study designby using VR. It thereby aims at establishing whether an ethical framework can describe decision-making in moral dilemma situations during car driving. This framework could then serve as a basis for algorithmsto be implemented in autonomous vehicles. Here we test five hypotheses: First, based on previous research we postulate that people would in general act in favor of the quantitative greater good, trying to keep the number of persons to be hit on a minimum level (Hypothesis 1). Yet, we speculate that the age of the potential victims matters in the sense that people might spare younger individuals as opposed to older ones (Hypothesis 2). In the traffic-specific context, we expect that pedestrians on the sidewalk would be protected, as they are not actively taking part in traffic. By staying on the sidewalk people generally, expect to be safe while implicitly giving consent to the finite risk of being injured when stepping on the street. Therefore, we hypothesize that people would avoid hitting pedestrians on the sidewalk in contrast to persons standing on the street (Hypothesis 3). On the other hand, we hypothesize that people prefer to protect children even if they are standing on the street as opposed to adults on the sidewalk (Hypothesis 4). Finally, we hypothesize that people would not reject self-sacrifice completely, but consider it when a high threshold of damage to others was reached (Hypothesis 5). To test these hypotheses, we implemented a driving simulation experiment with state-of-the-art virtual reality technologies following a study by Sütfeld et al. (2016). Participants were able to control a car as the driver and experienced various modified trolley dilemma situations as specified in the following.

## 2. Results

We analyzed data from 189 participants in a total of 4000 trials, distributed according to the hypotheses mentioned above into five modules. The results will be described in the following for each module separately.

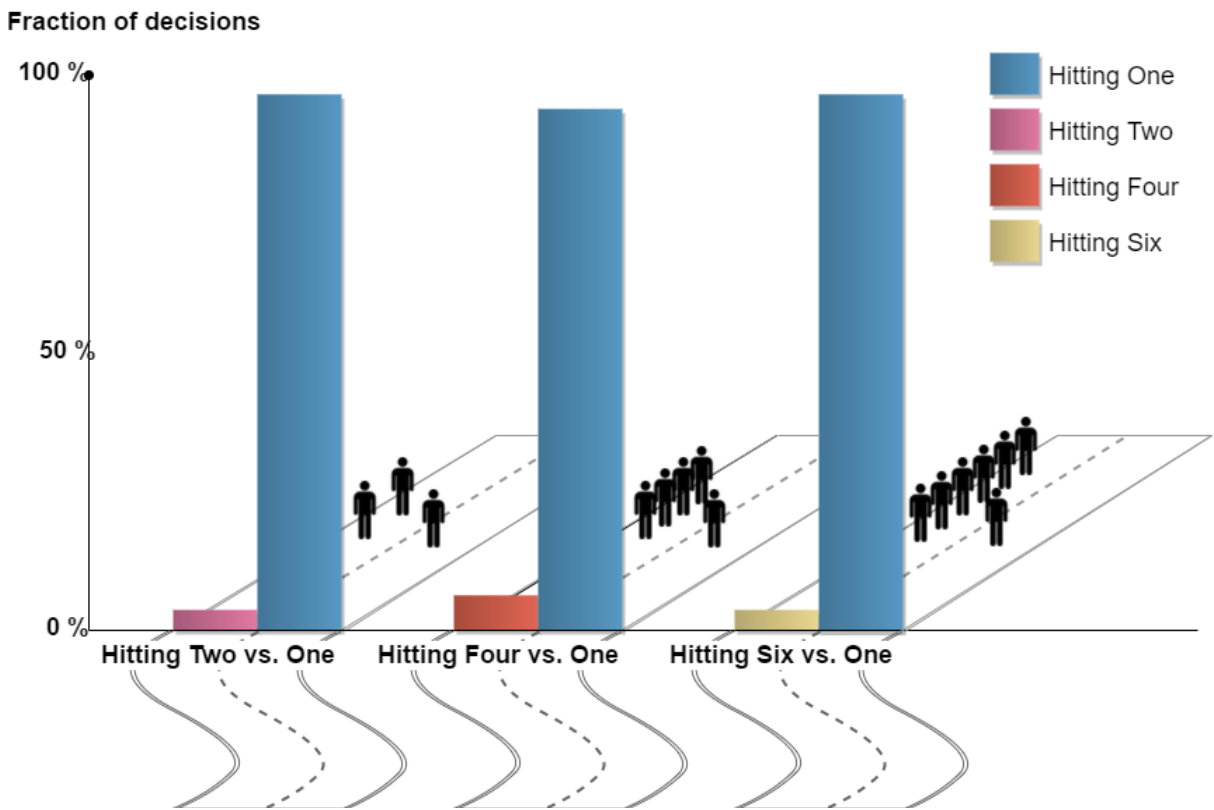

Figure 1: Decision distribution in the Quantitative Greater Good module. The graph depicts for every decision type in this module, how many participants decided one or the other way.

## 2.1 Quantitative Greater Good

In the first module, we tested whether people would act in favor of the quantitative greater good, by saving more instead of fewer avatars. This module consisted of three trials. The environment for this module was the suburban level setting, consisting of a two-lane road. We presented only standing adults as avatars. In the suburban level setting parked cars occupied both sides of the two-lane street. In the one versus two and one versus six conditions, only 7 out of 189 participants targeted the higher number of avatars (Figure 1). In the one versus four conditions, 12 participants targeted the four avatars instead of one. Thus, in all three conditions, the overwhelming majority of participants spared the larger number of avatars.

To investigate this difference between the conditions, we performed a permutation test. It yielded no significant difference ($p>0.05$).This shows that participants acted similarly throughout all three conditions. For each single condition the number of participants targeting one avatar instead of a larger number is highly significant ($p<0.01$). These data indicate that participants decided in favor of the quantitative greater good.

## 2.2 Age-Considering Greater Good

The second module tested the hypothesis that people would spare younger avatars in favor of older ones. It was composed of six trials in the suburban level setting. We used a child, an adult and an old man as avatars. Each trial presented one of the following six combinations of avatars: One child versus one standing adult, one child versus one old person, one standing adult versus one old person, one kneeling person versus one standing adult, one kneeling person versus one old person, and one kneeling person versus one child.

In the pairwise comparisons of children, adults and elderly we observed that the younger avatar was spared at the expense of the older avatar (Figure 2).A permutation test children vs. adults against elderly vs. adults yielded highly significant differences (p<0.001). This result demonstrates the inverse relation of the expected remaining lifespan of an avatar and the chance to get hit.Thisdecrease in value according to age was highly significant (p<0.01).

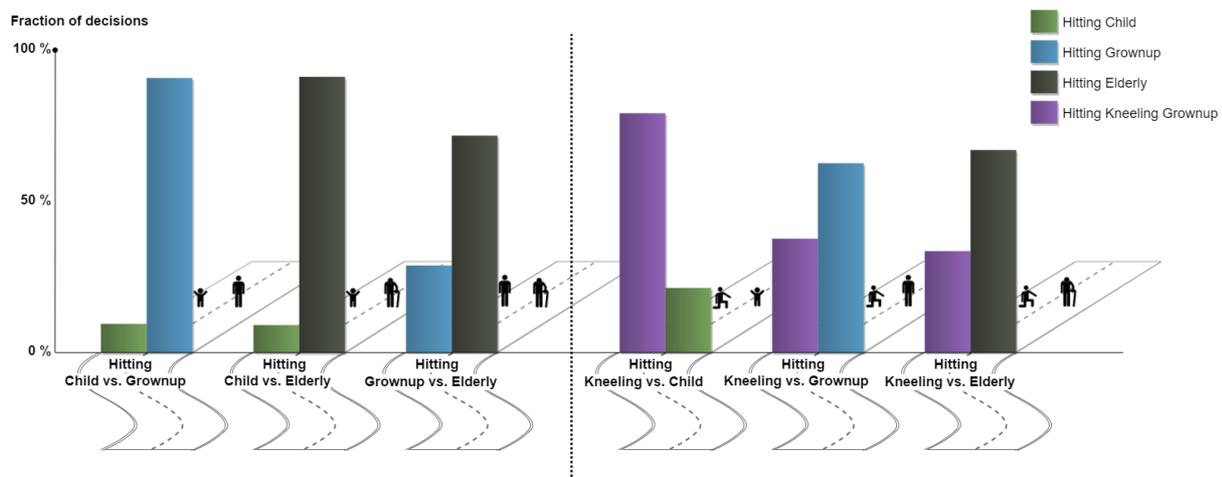

Figure 2: Decision distribution in module Age-Considering Greater Good. The graph depicts for every decision type in this module, how many percent of participants decided one or the other way. The left side shows purely age-considering decisions; the right side shows decisions about object height.

To investigate whether the difference emerged only through the variation in avatar height we tested kneeling adults vs. standing children, and elderly. We observed a highly significant difference in the children vs. kneeling adults comparison (Figure 2 fourth block, p<0.001). In the direct comparison of kneeling adult vs. standing adult, the latter was hit more often (p<0.001). A similar pattern emerged in the comparison kneeling adult vs. elderly. Thus, kneeling vs. standing moderates the participants' decisions to some degree. Yet, these results confirm that participants would spare younger avatars in favor of older ones irrespective of the height of the avatar.

## 2.3 The Influence of Context

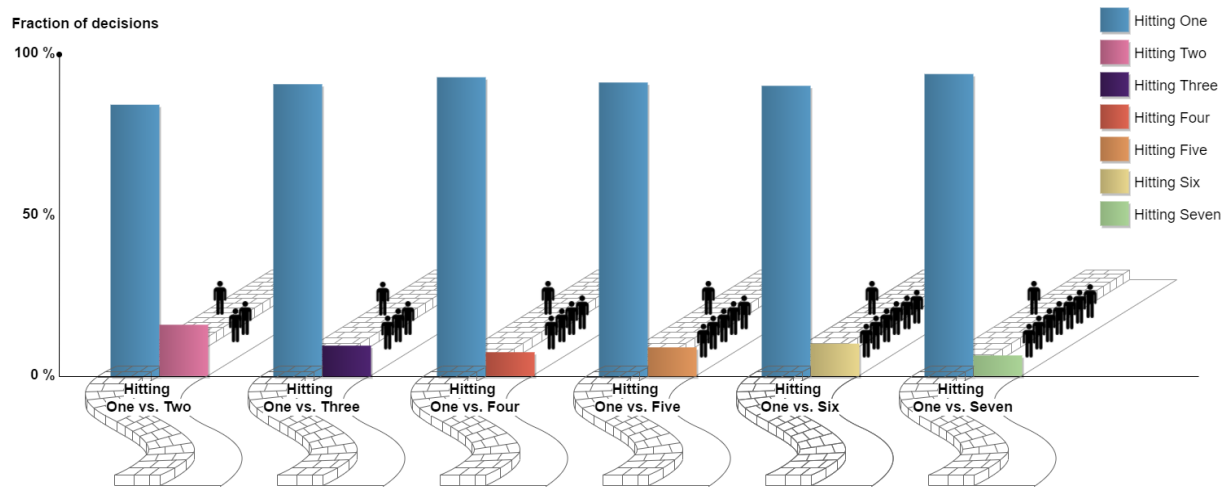

Figure 3: The Influence of Context module. The graph depicts for varying numbers of avatars the fraction of decisions sacrificing the single avatar on the sidewalk or the group of avatars on the street. The left lane is showing a sidewalk with the possibility to drive on, the right lane a one-way street.

In the third module, we explored the influence of context. Specifically, we hypothesized that avatars located on a sidewalk would be spared more often than on the street. Therefore, in direct analogy to the first module, we matched a single adult avatar on the sidewalk with two to six adult avatars on the street.

This module consisted of six trials in city levels using a two-lane street or a single lane plus sidewalk. These city level settings contained a one-way street with a sidewalk on both sides. One of the sidewalks was blocked by parked cars while participants had the opportunity to drive on the other one in order to avoid avatars on the street.

We hypothesized that, compared to the first module, a larger difference in the number of avatars would be necessary to lead to a consistent sacrifice of the single avatar on the sidewalk. Context did not seem to have a strong effect on decisions in general. Still, the majority of participants consistently sparedthe highest number of avatars possible, regardless of the sidewalk context (Figure 3). We investigated whether a switch point, defined by a critical imbalance of the number of avatars, could adequately describe the participants' decisions. That is if the number of avatars to be hit on the street is larger than this threshold participants change from driving on the street to driving on the sidewalk to save a large enough group of obstacles. Our data showed that only 2.56% of trials would need to be changed, for all participants to behave consistently according to a simple model with a single free parameter, the switch point.

For statistical evaluation, we fitted models, describing different switch points, to our data and compared the sums of squared residuals of the models, to identifythe model that fits the data best. Results showed that modeling the data with a switch point between the conditions with one versus two and one versus three avatars describes the data best (Figure 4), with a sum of squared residuals of 34.0. This, in turn, indicatesthat participants rather chose to drive on the sidewalk in order to save a group of three or more obstacles in contrast to saving only two obstacles. However, throughout all conditions, the number of participants driving on the sidewalk in order to save more obstacles is significantly higher than those trying to save the obstacle on the sidewalk.In comparison to the Quantitative Greater Good module, we only find minor quantitative differences. This shows that the sidewalk altogether shows a surprisingly small effect.

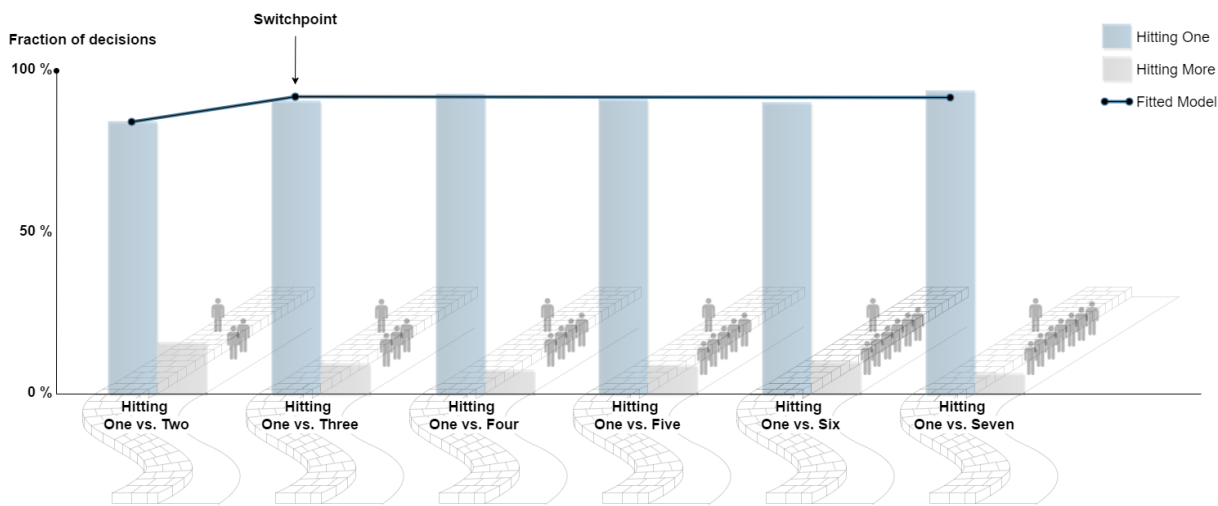

Figure 4: Depiction of the best-fitted model for decisions in the Influence of Context module. The raw data is depicted translucent, and the model is non-transparent. The model with the lowest sum of squared residuals had a switch point between two and three obstacles on the street.

## 2.4 Interaction of Age and Context

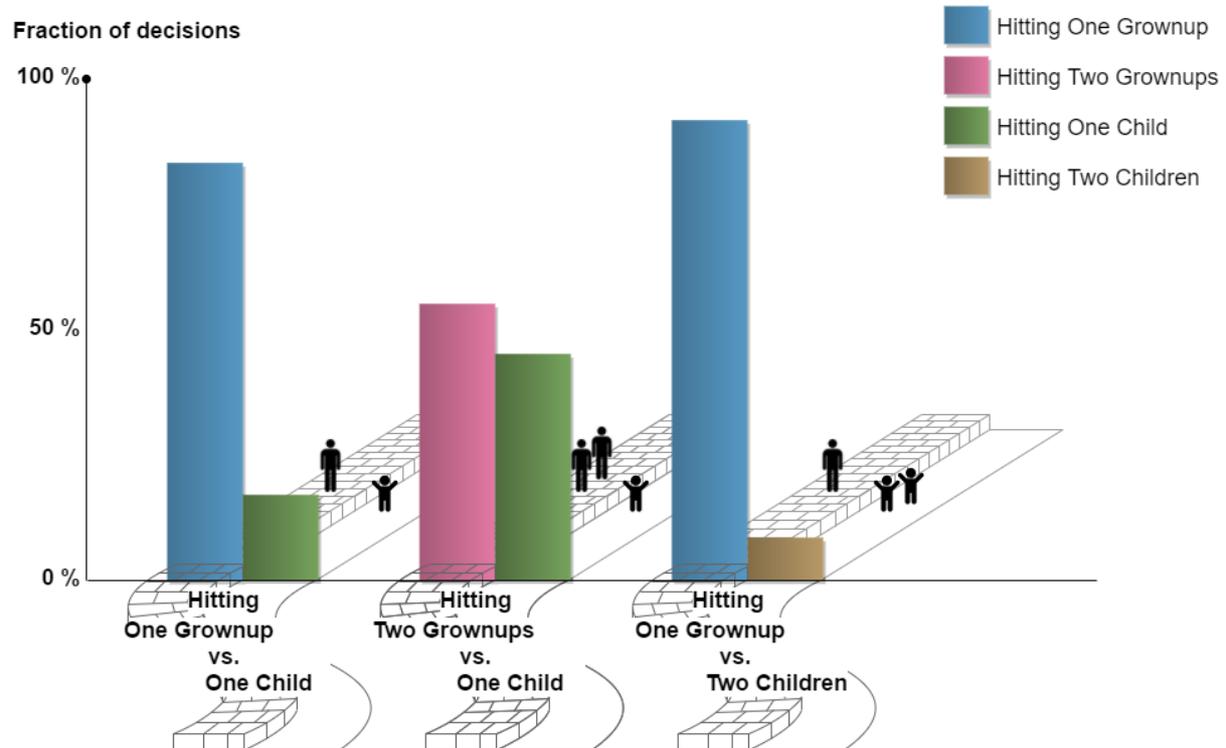

Figure 5: Interaction of Age and Context module. The graph depicts for one or two avatars of adults on the sidewalk and one or two avatars of children on the street the fraction of decisions sacrificing one or the other group. As in figure 3 and 4, the left lane is showing a sidewalk with the possibility to drive on. The right lane is a one-way street.

In the Age-Considering Greater Good and the Influence of Context modules, we had investigated the influence of age and context respectively in isolation. In the fourth module, we wanted to find out whether there is also an interaction of age and context. Hence, we also used the city level setting with the sidewalk and additionally included children avatars. There were three trials with the following combinations of avatars: two children on the street versus one adult on the sidewalk, one child on the street versus two adults on the sidewalk, and one child on the street versus one adult on the sidewalk.

Results showed that the majority of participants again spared children as opposed to adults despite the sidewalk context (Figure 5), as could be expected based on the findings from the previous modules.

In a further analysis, two permutation tests were performed to check for differences in the target actions of participants regarding the number of avatars. The conditions with one child on the street and one or two adults on the sidewalk were significantly different from one another ($p<0.001$). The same holds for the comparison of the condition with one child and one adult versus the condition of two children and one adult ($p<0.05$). The results were in accordance with the findings from all previous modules. Furthermore, the pattern of the results is compatible with independent effects of sidewalk and age.

## 2.5 Self-Sacrifice

In the fifth module, we investigated whether participants value their own life in the VR setup similarly to the value of the other avatars. That is, we gave them the possibility to save avatars at the price of sacrificing their own avatar. In close analogy to the previous modules, we investigated participants' choice as a function of the number of avatars in one of two groups. We hypothesized that the switch point, i.e. the number of avatars of the other group necessary to induce consistent decisions, would increase in comparison to the non-self-sacrifice condition in the first module.

The Self-Sacrifice module contained six trials in the mountain level 2, where we implemented a chasm on the right lane of the street with a construction sign in front of it. On the left lane, we presented a varying number of standing adults, ranging from two to seven avatars. The design was created to imply that participants would commit self-sacrifice within the experimental paradigm by driving off the cliff when driving in the right lane.

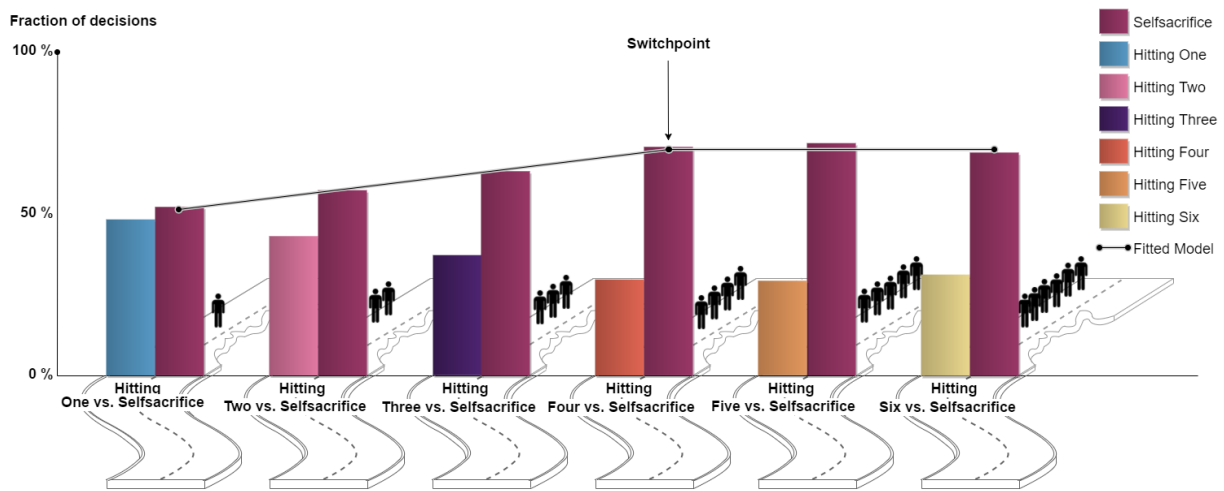

Figure 6: Self-Sacrifice module. The graph depicts the fraction of decisions for self-sacrifice or sacrifice a group of avatars. In each condition on the left lane is a normal street lane with a varying number of avatars. The right lane leads to a chasm. Continuing this way leads to a self-sacrifice. The best-fitted model is depicted as a line. The model with the lowest sum of squared residuals had a switch point between four and five obstacles on the street.

Analyzing the data of this module, we followed the same procedure as in the Influence of Context module. We postulated that a fixed threshold could describe the behavior of subjects. In case the number of avatars on the street was below the threshold, the group would be sacrificed. In contrast, when it was above the threshold subjects would choose self-sacrifice. We found that on average the decisions of only 5.2% of trials were not consistent with such a simple model.

To see if there is a general switching point, where people most likely change behavior towards self-sacrifice, we fitted a model via linear regression to the data. We fitted six models to the data, computed and compared the sums of squared residuals. The model with a switch point between the conditions with four and five obstacles best described our data (Figure 6). Then, with a value of 3, the sum of squared errors of this model is much smaller than the ones of the other models. This indicates that people are consistently willing to sacrifice themselves in case of being able to save a group of 5 or more obstacles with this decision.

However, it should be noted that about half of the participants were already consistently willing to sacrifice themselves to save only one avatar. The results of this module, therefore indicate that people are still acting in favor of the quantitative greater good even when their own life is at stake.

# 3 Discussion

Driving a car in virtual reality participants act in favor of the quantitative greater good. That is behavior consistently aimed at sparing as many avatars as possible at the expense of others. This even applied to situations in which participants had to virtually sacrifice their own avatar in order to save others. Age and context weakly modulated these behavioral patterns. Specifically we observed an inverse relation of the probability to sacrifice an avatar and the expected remaining lifespan of an avatar. Participants consistently saved younger avatars as opposed to older ones. The control condition using kneeling adults instead of children confirmed that this effect was not determined by the size of the visual appearance of the avatar. Surprisingly the context in the form of sidewalk versus street had only a small influence. Contrary to our expectations, the sidewalk did notsubstantially affect decision making, given that participants did not seem to be reluctant to drive on the sidewalk in order to spare a higher number of avatars.In conclusion, our results over all conditions support the hypothesis that people act in favor of the quantitative greater gooeven in scenarios involving a sidewalk or self-sacrifice.

Next, we will discuss potential limitations of the study. We phrased the dilemma in terms of sacrificing one or the other avatar. However, it is not self-evident that hitting one or more avatars would automatically lead to death. Indeed, some participants reported after the experiment, that the car was too slow to kill a human being. Before killing himself in the self-sacrifice trials, the participant had chosen to stop the car by driving into the group of avatars, risking injuries but no lethal damage. This might contribute to the equal treatment of groups larger than five people in the Self-Sacrifice module. On the other hand, phrasing the trolley dilemma, not in terms of life and death, but in terms of health or injury is of equal relevance.

The aspect that a collision might not necessarily lead to the death of the victim has to be considered when comparing avatars of different age. Potentially, participants were hitting adults more often instead of children not only because of age and expected lifespan, but also because they are less likely to die in case of a crash. It is not obvious, whether this argument is compatible with the preference to save children even when matched with kneeling adults. Due to the increased risk of fatal injury when hit by a car in a kneeling position, these avatars should have been treated similarly to children. Still, it is possible that participants were pondering more complex decision processes including severity of injury and risk of death. In social sciences, the term Disability-Adjusted Life Year(DALY) has been introduced (Murray, 1994). It is a complex measure and can be roughly understood as the number of years of a healthy life lost. Such a description would naturally explain the inverse relation between age and the probability of being spared. Thus, the decision process might be better described not by simply counting the number of lives, but as a more complex measure such as the DALY.

Moreover, general limitations of graphical display could have affected participants' decision processes and immersion. This appeals the fact that some participants dropped out of the experiment because they did not feel comfortable with virtually hitting avatars. This observation does not support a lack of realism or immersion. However, there seem to be many individual differences in play. In this regard, it cannot be ruled out for sure either that some, especially young, participants were not as committed to the study as expected but were mainly interested in the new VR technology offering a game-like experience. Thus, the average degree of immersion was rated high, but individual variations should be taken into consideration in further research addressing these problems.

In the field of implementing autonomous driving behavior, empiric knowledge is relatively sparse and ethical approaches are widely debated. Usable ADVs, as well as advanced simulation techniques like 3D virtual reality, are relatively new. Consequently, empirical studies rely heavily on questionnaires directing issues straight at the potential customers. The behavior of ADVs and their control algorithms will be judged by the standards and ethics of the society in which they operate. This again emphasizes the crucial role of acceptance. It seems reasonable since self-driving cars need moral algorithms capable of three aspects: being consistent, not causing public outrage and not discouraging potential buyers (Bonnefon, Shariff, & Rahwan, 2015). Since it is arguable whether artificial agents can truly exhibit moral behavior, the programming behind these cars will be judged. Therefore, the problem is reflected back to humans. ADVs not only have to embody the laws, but also the ethical principles of the society they operate in (Gerdes & Thornton, 2015).

Various studies were conducted, given the assumption that passengers would like ADVs to behave similarly to humans (Goodall, 2014; Malle, Scheutz, Arnold, Voiklis, & Cusimano, 2015; Sikkenk & Terken, 2015). They found that many factors drastically influence human behavior in traffic, e.g. weather conditions and the driving style of other traffic participants. This raises concerns about a uniform behavior in ADVs(Sikkenk & Terken, 2015). Such differences do not only occur in driving but also in judging decisions of humans and machines. Another study examined the differences in responsibility between humans and machines in cases of an inevitable fatal crash. Participants had to judge the decision of either a human driver in a dilemma situation, or an autonomous car deciding on its own. In contrast to human drivers, where utilitarian decisions were most favorable, participants expected ADVs to behave in a utilitarian manner under all circumstances (Li, Zhao, Cho, Ju, & Malle, 2016; Malle et al., 2015).

These studies point out, that the general population seems to favor utilitarian decisions(Bonnefon et al., 2015; Li et al., 2016; Malle et al., 2015). This applies even to cases, where the driver has to sacrifice himself for the greater good(Sachdeva, Iliev, Ekhtiari, & Dehghani, 2015). Such behavior is in line with the general philosophers' opinion(Fischer & Ravizza, 1992) and can be understood as an act of maximizing utility (Thomson, 1985). Therefore, it is mostly referred to as utilitarian reasoning and decision-making. On first thought, utilitarian decisions offer themselves to a quantitative treatment and appear to be suitable for ADVs. However, the problem of how to implement ethics in machines, especially in dilemma situations remains. A recent project at the Bristol robot laboratories (Winfield, Blum, & Liu, 2014) showed, that there is no such thing as a simple rule, like the first Asimov robot law, to save human life when it comes to dilemma situations. To fix that, more rules have to be applied to the code. On matters of judgments like these, not even humans always agree (Deng, 2015). These seem to be crucial aspects to promote ADVs to potential customers and allow them to be an integral part of our society.

In general, it is questionable whether machines should mimic human behavior just to encourage potential customers given that this might entail a naturalistic fallacy. To avoid this case, other articles argue against the moral Turing test as framework for moral machine performance, since it is vulnerable to deception, inadequate reasoning and inferior moral performance. The moral Turing test, in form of obligations and norms, does not settle morality of an action fully. Furthermore,when it comes to a fatal crash the utilitarian approach is forbidden by law. To protect the human dignity of each individual, a classification of value, meaning discrimination, is not allowed according to §1 Abs 1, GG and the 14th amendment of the US constitution(Lin, 2015). A transparent, accountable process of reasoning, reliably prefiguring moral performance in line with current law is suggested instead.Autonomous systems should be predictable, controlled and transparent - allowing explicit reworking and recasting (Arnold & Scheutz, 2016). In conclusion, rather than hard-and-fast rules now, review boards are suggested which would provide a process to allow manufacturers, lawyers, ethicists, and government entities to work through these ethical decisions (Kirkpatrick, 2015).

Another option to potentially deal with moral issues would be to give control back to the driver during periods of congestion or treacherous conditions just like the German government suggests(Hevelke & Nida-Rümelin, 2014). In this case the machine is not fully autonomous and an accountable driver would still be required so that the ADV wouldn't show full potential(Kirkpatrick, 2015). However, as such crucial situations are usually associated with severe time constraints, giving control back to the driver might quickly amount to no decision. That this course of action, however, is optimal or desirable is questionable.

As a further alternative, a form of a hybrid between the moral Turing test and a straight logical approach has been suggested. Because ethical dilemmas do not necessarily have objective answers, but significant ethical implications for the user, key elements for solving the question how ADVs should behave are therefore ethicists as experts for ethical evaluations of robotics(Millar, 2016).

Despite manyunsolved issues, fully autonomous self-driving cars would improve mobility for elderly or disabled people, reduce crashes, decimate annual fatalities in traffic, ease congestion, improve fuel economy, reduce parking, and offer mobility to those unable to drive. The U.S. economic benefits could reach around 25 billion dollar per year with only 10% market penetration. Including high penetration rates, this raises the annual benefit up to $430 billion, which makes ADV a technology of a better future(Fagnant & Kockelman, 2015). The number of avoided fatalities is a sufficient reason to promote ADVs. Therefore, the idea of McBride(2016)for partial automation only is decidedly rejected here. Instead, further research addressing open questions should be encouraged. These range from technical issues, to ethical and psychological problems, as well as legal aspects like responsibility and policy issues.

Summarizing, our results show that participants behave consistently utilitarian in various dilemma situations. These decisions were only slightly modulated by context, such as a sidewalk. Further, the effect of age might be subsumed in a utilitarian decision process as well. Even in conditions involving a self-sacrifice participants decisions were compatible with a utilitarian strategy. Such strong behavioral tendencies must be considered in the implementation of moral decisions in ADVs especially when aiming at a high acceptance in society.

In the discussion of a broadly applicable ethical framework, our research explores the foundation for such a framework. The study provides a basis for an algorithm implementing morals in ADVs by describing how human car drivers would behave in these conditions and therefore what is seen fit as behavior in general traffic situations.

# 4 Material and Methods

## 4.1 Participants

216 unpaid subjects participated in the study. Participants were acquired from various venues. 142 were tested in the laboratory or cafeteria and at the alumni conference of Cognitive Science of the University of Osnabrück. 74 participants were tested at public locations, such as the waiting rooms of the City Hall of Osnabrück and at a local inspection authority, the so-called KFZ-Prüfstelle. Data from 27 participants had to be excluded from the analysis for various reasons: 15 participants did not complete the experiment due to nausea or disagreement with the experimental settings. Moreover, 12 participants were excluded from analysis as they had failed more than six times in the training trials. In the end data from 189 participants served for analysis (62 female, 127 male). They were aged between 18 and 67 years with a mean of 24.32 years.

## 4.2 Apparatus and Setup

For the technical apparatus at every venue,we used a computer equipped with an Intel Xeon E5620 with 2.4 GHz, 12 GB RAM, and the 64-bit operating system Windows 7 Professional. The NVIDIA GeForce GTX 970 and NVIDIA Quadro NVS 295 (GPU) served for graphics. We used the Oculus Rift DK2 in combination with Bose Noise Cancelling Headphones as VR equipment.

## 4.3 Stimuli and Design

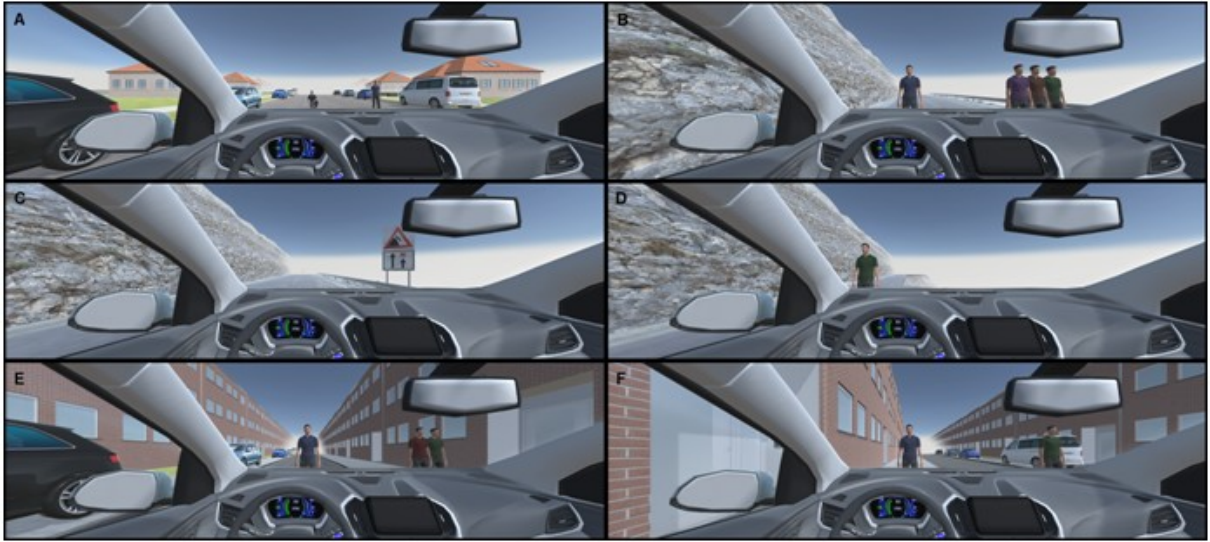

The experiment was designed as a computer application with featured movement in a virtual reality.Figure 7: Screenshots of the experiment virtual reality environment. The participants could decide between driving on one of two lanes and thus decide which of two avatars to hit or hit an avatar vs. driving into a chasm. The screenshots are taken out of the modules: A: Age-considering Greater Good in the suburban setting. B: Quantitative Greater Good module in the mountain setting. C: Self-Sacrifice module showing the road sign warning of the oncoming chasm in the mountain setting. D: Self-Sacrifice module in the mountain setting. E: Age-considering Greater Good module in the city setting with cars on the left side. F: Age-Considering Greater Good module in the city setting with cars on the right side.

One training track setting and five different experiment track settings were presented. In each setting, participants were driving a car on a two-path track. The car was driving at a constant speed of 36 km/h, which was displayed on a screen in VR, and the tracks ranged between 180m and 200m to avoid habituation to the trial length. The surroundings varied between five different environmental settings - a suburban, two mountain and two city level settings. The starting lane (left/right lane) was randomized for each trial, only in the city level participants always started on the drivable lane to avoid the sidewalk as starting position.

In order to decrease the visual range and thereby guarantee a constant decision-time of four seconds, all settings included foggy weather. A beeping sound indicated to the participants that they had control over the vehicle. The relatively low speed of the car was selected as a compromise to allow a reasonable time for deliberation and yet have the nature of the obstacle clearly visible. At 15 meters distance from the avatars another beeping sound signaled that the control over the vehicle was withdrawn as later inputs would have led to incomplete lane change maneuvers.

The presented avatars were only male to avoid an effect of gender difference. As previous studies showed that male and female avatars are treated differently, we decided to use only males in order to avoid an effect of gender difference (Sütfeld et al., 2016). These distinct types of avatars appeared on the lanes in different combinations as specified in the following:

The *Quantitative Greater Good module* consisted of three trials to test Hypothesis 1. The environment for this module was the suburban and the mountain level setting, both consisting of a two-lane road. We presented only standing adults presented as avatars. In the suburban level setting parked cars occupied both sides of the two-lane street. The setting included a suburban style environment with houses and front lawns. There was always one avatar on one laneas opposed to either two, four or six avatars on the other lane. The respective sides were randomized. The mountain level setting consisted of a mountainside as limitation on the left and a crash barrier in front of a steep ravine on the right side of the two-lane street.

The *Age-Considering Greater Good module* aimed at testing Hypothesis 2. It was composed of six trials in the suburban level setting. We used a child, an adult and an old man as avatars; additionally a kneeling adult served to make sure that possible effects were not only due to the size of the stimuli, given that the kneeling adult was of the same height as the child. Each trial presented one of the following six combinations of avatars: One child versus one standing adult, one child versus one old person, one standing adult versus one old person, one kneeling person versus one standing adult, one kneeling person versus one old person, and one kneeling person versus one child.

The *Influence of Context module* consisted of six trials in the city levels using a two-lane street or, in the investigation of hypothesis 3, a single line plus sidewalk. These latter city level settings contained a one-way street with a sidewalk on both sides. Parked cars blocked one of the sidewalks, while participants had the opportunity to drive on the other one in order to avoid avatars on the street. In city level setting 1, the sidewalk to drive on was on the left side whilst in city level setting 2 the sidewalk to drive on was on the right side of the lane. We randomized the number of trials within city level setting 1 or city level setting 2 between subjects. There was always one standing adult on the sidewalk and a number of standing adults on the street, varying between two and seven.

The *Interaction of Age and Context module* testing Hypothesis 4 was based on the *Influence of Context module* also using the city level settings, including children as avatars. The trial settings were also intersubjectively randomized. There were three trials with the following combination of avatars: two children on the street versus one standing adult on the sidewalk, one child on the street versus two standing adults on the sidewalk, and one child on the street versus one standing adult on the sidewalk.

The *Self-Sacrifice module* investigated Hypothesis 5 contained six trials in the mountain level 2. Here, we implemented a chasm on the right lane of the street with a construction sign in front of it. A varying number of standing adults, ranging from two to seven avatars, was presented on the left lane. On the right side, a cliff and barricade with the corresponding street sign appeared. These avatars were created to imply that participants would commit self-sacrifice within the experimental paradigm by driving over the cliff when driving on the right lane.

## 4.4 Procedure

Upon arrival, we informed participants about the content and the procedure of the experiment and asked them to sign a consent form clarifying that they were able to terminate the experiment at any time without stating any reason. We informed them that their data would be saved anonymously. Additionally, they were asked whether they had experienced a traumatic car accident previously. If this question had been positively answered, the participant would have been excluded from the experiment. Once the consent was given, participants were seated at a desk and provided with predefined oral instructions. They were told that the VR experiment consisted of three phases: training trials, experimental trials and a questionnaire. During the trials, they would be alone in the car as the driver and would drive always on one of two lanes in different environmental levels. Participants were furthermore instructed to use the left or right arrow keys to change the driving lane in the trials and to additionally press the spacebar to confirm their answers in the instruction and questionnaire sections. They were also informed that the monitor for overviewing during the instructions would be turned off while performing the experimental trials and answering the questionnaire. Respective written instructions were additionally presented within the experiment. Finally, we adjusted the Oculus Rift as well as the headphones to their head and the experiment started. Three initial training trials served to get accustomed to the VR environment and to obtain control over the car. To pass the training phase, in each training trial participants had to avoid three pylons appearing on one of the lanes alternately with the first one always on the starting lane. If they hit a pylon, the trial had to be repeated. After successfully completing all three training trials, the various types of avatars to hit or to spare in the experimental trials were presented one by one and then the trials started. In the end, participants answered the questionnaire to finish the experiment. The questionnaire data is, however, beyond the scope of the present paper and will be presented elsewhere. The duration of the whole experiment was approximately 15-20 minutes.

## 4.5 Statistical tests

For all analyses, we used Python 2.7. Only choices for the left or right lane and the avataron the lane finally selected were taken into account. After descriptively analyzing the data, statistical tests were performed, to check our hypotheses and investigate the significance of the results.

To test, if participants choose a smaller number of avatars to spare a larger number in inevitable crash scenario, we performed a permutation test on the complete data set of the conditions of Quantitative Greater Good module, investigating the influences of the number of avatars on participants' decisions. This accounts also for the Self-Sacrifice, Age-Considering Greater Good and Influence of Context module for each condition individually. Additionally, the significance of a binomial test using pooled data proved significance in comparison to our null-hypothesis of a random distribution of choices in the aforementioned trials.

For the Age-Considering Greater Good module, as well as for the Self-Sacrifice module, we calculated fractions of trials that would need to be changed in order for each participant to act consistently in their decisions. This error rate would naturally be around 5% (Kuss, Jäkel, & Wichmann, 2005). In order to test whether our data fits the hypothesis of a switch point, we fitted different models on the data and computed the performance of each. Sixmodels were fitted, each based on another underlying switch point. As we assumed that upon a certain switch point participants would not switch back, we calculated the mean between the conditions with an avatar number higher than a certain switch point and assumed that the model would pass through this mean in a plateau. For the conditions with avatar numbers smaller than the underlying switch point, we assumed a linear increase up to the calculated mean. To test which model fits the data best, we computed and compared the sums of squared residuals. After that, a permutation test was applied to check for inconsistencies in decisions with more avatars than our assumed switching point for both modules.

# 5 Acknowledgements

The authors like to thank all participants in the study project: AaliaNosheen, Max Räuker, Juhee Jang, Simeon Kraev, Carmen Meixner, Lasse T. Bergmann and Larissa Schlicht. This study is complimented by a philosophical study with much broader scope (Lariss Schlicht, Carmen Meixner, Lasse Bergmann). We gratefully acknowledge the support by H2020 – H2020-FETPROACT-2014 641321 – socSMCsproject.

# 6 Financial Interests

This publication presents part of the results of the study project "Moral decisions in the interaction of humans and a car driving assistant." Such study projects are an obligatory component of the master degree in cognitive science at the University of Osnabrück. It was supervised by Peter König, Gordon Pipa, and Achim Stephan. Funders had no role in study design, data collection, and analysis, decision to publish, or preparation of the manuscript.

# 7 Author contributions

This study was planned and conducted in an interdisciplinary study project supervised by Prof. Dr. Peter König, Prof. Dr. Gordon Pipa, and Prof. Dr. Achim Stephan. Maximilian Alexander Wächter, Anja Faulhaber, and Silja Timm shaped the experimental design to a large degree. Leon René Sütfeld had a leading role in the implementation of the virtual reality study design in Unity. Anke Dittmer and Felix Blind contributed to the implementation. Anke Dittmer, Felix Blind, Silja Timm and Maximilian Alexander Wächter contributed to the data acquisition, analysis and writing process. Anja Faulhaber contributed in the data acquisition and the writing process.

*Poster Presented at KVIT 2016, Linköping, SE*.

Thomson, J. J. (1976). Killing, letting die, and the trolley problem. *The Monist*, *59*(2), 204–217. https://doi.org/10.5840/monist197659224

Thomson, J. J. (1985). The Trolley Problem. *The Yale Law Journal*, *94*(6), 1395. https://doi.org/10.2307/796133

Unger, P. (1996). *Living high and letting die: Our illusion of innocence*. Philosophy and Phenomenological Research. https://doi.org/10.1093/0195108590.001.0001

Winfield, A. F. T., Blum, C., & Liu, W. (2014). Towards an Ethical Robot: Internal Models, Consequences and Ethical Action Selection. In *Advances in Autonomous Robotics Systems* (pp. 85–96). Springer, Cham. https://doi.org/10.1007/978-3-319-10401-0_8